# Comment: On Random Scan Gibbs Samplers

**Richard A. Levine and George Casella**

## 1. INTRODUCTION

We congratulate the authors on a review of convergence rates for Gibbs sampling routines. Their combined work on studying convergence rates via orthogonal polynomials in the present paper under discussion (which we will denote as DKSC from here onward), via coupling in Diaconis, Khare and Saloff-Coste (2006), and for multivariate samplers in Khare and Zhou (2008), enhances the toolbox of theoretical convergence analysis. This has the potential of opening new avenues of pursuit for gauging chain convergence in practice, and optimally implementing Gibbs sampler strategies. In this discussion, we focus on the latter, within the context of the random scan Gibbs sampler presented in DKSC. Although the analysis in DKSC does not seem to extend to the random scan implementation we consider, a study of convergence rate and estimator precision is possible, in theory, for special cases as well as in general practice. Our aim is to motivate further research within the context of DKSC to identify objective criteria for optimizing implementation of the random scan Gibbs sampler.

## 2. REVISITING RANDOM SCAN GIBBS SAMPLERS

The random scan Gibbs sampler considered in DKSC has an equal likelihood of visiting each coordinate, $(x, \theta)$, during an iteration of the sampler.

*Richard A. Levine is Associate Professor, Department of Mathematics and Statistics, San Diego State University, San Diego, CA 92182, USA e-mail:*
ralevine@sciences.sdsu.edu. *George Casella is Distinguished Professor, Department of Statistics, University of Florida, Gainesville, FL 32611, USA e-mail:* casella@stat.ufl.edu.

As put forth by the seminal convergence theory work of Liu, Wong and Kong (1995) and discussed more recently by Levine and Casella (2006), an optimal implementation of the random scan strategy may visit less often components with a marginal that is easier to understand or describe. For example, in the bivariate cases of DKSC, each iteration of the random scan visits $x$ with probability $\alpha_1$ and $\theta$ with probability $1 - \alpha_1$, where $\alpha_1 \in (0, 1)$, not necessarily equal to 0.5. For the general multivariate problem of sampling a $d$-vector $\mathbf{X}$, the random sweep strategy is characterized by selection probabilities $\boldsymbol{\alpha} = (\alpha_1, \alpha_2, \ldots, \alpha_d)$, where $\sum_{i=1}^{d} \alpha_i = 1$, $\alpha_i$ not necessarily equal to $1/d$ for all $i$.

In the notation of DKSC, the transition kernel of the random scan Gibbs sampler for a function $g \in L^2(P)$ is

$$
\begin{aligned}
\bar{K}g(x, \theta) = {} & \alpha_1 \int_{\Theta} g(x, \theta')\pi(\theta'|x)\pi(d\theta') \\
& + (1 - \alpha_1) \int_{\mathcal{X}} g(x', \theta)f_\theta(x')\mu(dx').
\end{aligned}
\tag{1}
$$

Unfortunately, $\bar{K}$ in (1) is not readily diagonalizable as the decomposition in the proof of DKSC Theorem 3.1, part (c), relies on the equal selection probabilities ($\alpha_1 = 0.5$) to partition the transition kernel acting on appropriate functions $g$. However, in the cases of discrete state spaces and Gaussian target distributions, both considered in the exposition of DKSC, we may identify explicit convergence rates and optimally choose selection probabilities. In the following sections, we elaborate on these findings and present an alternative approach with estimator precision as an objective criterion. We also suggest avenues for future research within the context of DKSC to address the random scan Gibbs sampler decision problem.

## 3. CONVERGENCE RATES

Convergence rates of Gibbs sampling routines may be formulated in two special cases: Gaussian and discrete target distributions. DKSC Section 6.3 eludes to the case of Gaussian distributions, identifying the work of Goodman and Sokal (1989), that shows convergence rates as the largest eigenvalue of a matrix related to the dispersion matrix and an autoregressive transition of the Markov chain (see Khare and Zhou (2008), as well). Amit (1996) and Roberts and Sahu (2001) provide an alternative expression which lends well to our analysis of random scan Gibbs samplers. In particular, Levine et al. (2005) shows that for a $d$-dimensional Gaussian target distribution with $d$-vector zero mean and dispersion matrix $\mathbf{\Sigma}$, $N_d(\mathbf{0}, \Sigma)$, the random scan Gibbs sampler has convergence rate $\rho(I - \mathbf{\Psi SR})$ where $\rho(\cdot)$ is the spectral radius (maximum modulus eigenvalue), $\mathbf{\Psi} = \text{diag}(\alpha_1, \dots, \alpha_d)$, $\mathbf{R} = \mathbf{\Sigma}^{-1}$ and $\mathbf{S} = \text{diag}(1/r_{11}, \dots, 1/r_{dd})$ with $r_{ii}$ the $(i, i)$th or $i$th diagonal element of $\mathbf{R}$. Note that $\rho(I - \mathbf{\Psi SR})$ is a function of the selection probabilities and thus may be used as an objective criterion for optimal choice of $\boldsymbol{\alpha}$.

Consider a different form of the Gaussian example of DKSC Section 4.3, along the lines of Section 6.3, a bivariate Gaussian target distribution with bivariate mean of zero, standard deviations $\sigma_1$ and $\sigma_2$, and correlation $\rho$. The convergence rate is

$$\lambda_{rs} = 0.5\{1 + \sqrt{1 + 4\alpha_1^2(1 - \rho^2) - 4\alpha_1(1 - \rho^2)}\}.$$

Interestingly, the covariance structure, with covariance $\tau = \rho\sigma_1\sigma_2$, leaves the convergence rate as a function of the correlation $\rho$ and not the variance components. The random scan with equal selection probabilities, $\alpha_1 = 0.5$, has the smallest convergence rate, over the range of standard deviations and correlation.

In the case of multivariate Gaussian distributions, the random scan with equal selection probabilities is not necessarily optimal with respect to convergence rate. For example, consider a trivariate Gaussian distribution with zero mean vector and dispersion matrix

$$\mathbf{\Sigma} = \text{diag}(\sigma_1^2, \sigma_2^2, \sigma_3^2) - 1/(d + 0.005)\mathbf{J}$$

where $\mathbf{J}$ is a matrix of ones, an exchangeable correlation structure considered by Roberts and Sahu (2001) and Levine et al. (2005). In the case $\sigma_1 = 10$ and $\sigma_2 = \sigma_3 = 1$, the random scan Gibbs sampler with $\boldsymbol{\alpha} = (0.22, 0.39, 0.39)$ has the smallest convergence rate. Nonetheless, the gain in rate over the random scan with equal selection probabilities is less than 10%. Levine et al. (2005) provide further illustrations of optimal random scan Gibbs samplers for multivariate Gaussian target distributions where non-equal selection probabilities minimize the convergence rate. However, often the computational cost in identifying the selection probabilities that minimize the convergence rate is not sufficiently offset by the gain in convergence speed.

In the case of discrete state spaces, Frigessi et al. (1993) shows that for a transition matrix $P_{rs}$, the convergence rate is the second largest eigenvalue in modulus, $\rho_2(P_{rs}) = \max\{|\lambda| : \lambda$ an eigenvalue of $P_{rs}$, $\lambda \neq 1\}$, the largest eigenvalue being equal to one. Note that $\rho_2(P_{rs})$ is a function of the selection probabilities and thus may be used as an objective criterion for optimal choice of $\boldsymbol{\alpha}$.

Consider the binomial example of DKSC Section 5.1 where at iteration $t$, $\theta|X_{t-1} \sim$ hypergeometric$(n_1, n_2, X_{t-1})$, $X = \theta_{t-1} + \epsilon$ with $\epsilon \sim$ binomial$(n_2, p)$ and marginally $X \sim$ binomial$(n_1 + n_2, p)$, $\theta \sim$ binomial$(n_1, p)$. Of course the cardinality of the state space is a function of $n_1$ and $n_2$ so a closed form expression of the convergence rate as a function of $\alpha_1$ is not available. However, for given $n_1$ and $n_2$, we may easily minimize $\rho_2(P_{rs})$ with respect to the selection probabilities. Empirical evidence suggests that the random scan with equal selection probabilities, $\alpha_1 = 0.5$, leads to the smallest convergence rate.

Levine (2005) provides illustrations of optimal random scan Gibbs sampler for multivariate discrete target distributions. As with multivariate Gaussian target distributions, the random scan with equal selection probabilities is non-optimal with respect to convergence rate, however the loss in convergence speed is minimal. The random scan Gibbs sampler analyses in DKSC for bivariate chains, and that of Khare and Zhou (2008) for multivariate target distributions, may thus be a worthwhile pursuit, focusing exclusively on uniform visitation of coordinates. We will discuss this matter more below.

## 4. ASYMPTOTIC VARIANCE

We have seen that if the optimality criterion is convergence rate, it is often the case that there is only minimal gain in using the optimal random scan rather than the equal probability scan. The story is not the same, however, if we shift attention to estimator precision as the objective criterion.

An alternative means of choosing random scan selection probabilities is through a study of estimator precision. Suppose interest lies in estimating $E_\pi\{h(\mathbf{X})\}$ for a function $h \in L^2(\pi)$, where $\pi$ is the distribution of the $d$-vector $\mathbf{X}$. The natural estimator of this expected value is the sample mean, $(1/m)\sum_{i=1}^{m} h(\mathbf{X}_i)$ of $m$ variates generated by the random scan Gibbs sampler. We may thus identify the best scan strategy through a minimization of the asymptotic variance

$$(2) \quad R(\boldsymbol{\alpha}, h) = \lim_{m \to \infty} m \operatorname{VAR}\left\{\frac{1}{m}\sum_{i=1}^{m} h(\mathbf{X}_i)\right\}.$$

Levine and Casella (2006) show that the two-lag autocovariance in the asymptotic variance expansion may be presented as the square of the convergence rate, relating these two objective criteria. (See Chen, Liu and Wang (2002), for more details on this relationship.) Levine and Casella (2006) also show that $R(\boldsymbol{\alpha}, h)$ is a polynomial in $\boldsymbol{\alpha}$. For sake of space, we will not duplicate the expressions here. However, optimization of the asymptotic variance over the selection probabilities is feasible, particularly in the case of Gaussian and discrete target distributions.

Consider again the bivariate Gaussian example (DKSC Sections 4.3 and 6.3). A second-order approximation of the asymptotic variance (2) for linear functions $h(\mathbf{X})$ identifies optimal random scans with non-equal selection probabilities, following the intuition presented earlier of visiting more often the most variable coordinate. For example, in the case of estimating the sum of the coordinates, the asymptotic variance is

$$\begin{aligned}
R(\boldsymbol{\alpha}, h) = {} & \sigma_1^2 + \sigma_2^2 + 2\rho\sigma_1\sigma_2 + \alpha_1(\rho\sigma_1 + \sigma_2)^2 \\
& + (1 - \alpha_1)(\sigma_1 + \rho\sigma_2)^2 + \alpha_1^2(\rho\sigma_1 + \sigma_2)^2 \\
& + (1 - \alpha_1)^2(\sigma_1 + \rho\sigma_2)^2 \\
& + 2\alpha_1(1 - \alpha_1)(\rho\sigma_1 + \sigma_2)(\sigma_1 + \rho\sigma_2)\rho.
\end{aligned}$$

If the standard deviations are $\sigma_1 = 2$ and $\sigma_2 = 1$ with correlation $\rho = 0.5$, the scan that minimizes the asymptotic variance has $\alpha_1 = 0.93$.

In the case of discrete state spaces, Peskun (1973) shows that the asymptotic variance is $R(\boldsymbol{\alpha}, h) = h(2\mathbf{B}\mathbf{Z} - \mathbf{B} - \mathbf{B}\mathbf{A})h^T$ where $\mathbf{A}$ is a matrix with each row containing the vector of stationary distribution probabilities $\pi$, $\mathbf{B}$ is a diagonal matrix with $\pi$ on the diagonal, $\mathbf{h}$ is a vector of the function $h$ applied to each element of the state space, and $\mathbf{Z} = \{\mathbf{I} - (\mathbf{P}_{rs} - \mathbf{A})\}^{-1}$ the fundamental matrix with identity matrix $\mathbf{I}$. Consider again the binomial example of DKSC Section 5.1. As in the Gaussian case, minimization of this asymptotic risk over $\alpha_1$ identifies optimal random scans that visit the most variable coordinate at a higher frequency. For example, if parameters are set at $n_1 = 6$, $n_2 = 3$, and $p = 0.5$, the scan that minimizes the asymptotic variance has $\alpha = 0.56$.

Levine et al. (2005) and Levine and Casella (2006) show that these optimal random scans present significant improvement over a random scan with equal selection probabilities, not only in asymptotic variance but also in chain mixing.

## 5. IMPLEMENTATIONS

For general applications of the random scan Gibbs sampler to multivariate target distributions, neither the convergence rate nor the asymptotic variance may necessarily be available in closed form. However two implementations have been proposed to choose optimal selection probabilities in practice. Levine et al. (2005) suggests using a Gaussian approximation to the target distribution to determine the optimal random scan, perhaps in a tuning phase of the sampler or as an adaptive procedure. Since the convergence rate and asymptotic variance are accessible under a Gaussian target distribution, several adaptive and non-adaptive random scan Gibbs sampler algorithms present themselves.

Levine and Casella (2006) propose an adaptive random scan Gibbs sampler which chooses optimal selection probabilities "on the fly," learning and adapting the sweep strategy as the chain traverses the state space. The induced chain is no longer Markov but still converges to the desired equilibrium distribution. In the most general form, the adaptive strategy identifies a minimax random scan for the set of selection probabilities that minimizes the asymptotic variance for the worst possible function of interest.

The optimal random scan Gibbs samplers determined with respect to the convergence rate and asymptotic variance, though potentially identifying different sets of selection probabilities, are not contradictory. As suggested by Mira (2001) and discussed further in Levine et al. (2005), the convergence rate criterion is most desirable during the burn-in period of the Markov chain, estimator precision is of importance for drawing inferences from the Gibbs sampler output. Therefore, our recommendation is to first implement a random scan with equal selection probabilities and then, during the post-processing phase of the sampler, choose selection probabilities that minimize the asymptotic variance. The convergence rate calculations of DKSC are of utmost importance then for convergence assessment. If the techniques allow for matrix decompositions or expressions for the asymptotic variance, the tools provide for pre- and post-burn-in implementations of the random scan Gibbs sampler. Furthermore, such expressions may lend to computationally inexpensive implementations of both the Gaussian approximation and adaptive procedures for optimally selecting random scan Gibbs samplers.

## ACKNOWLEDGMENTS

Supported by NSF Grants DMS-04-05543, DMS-06-31632 and SES-0631588.

## REFERENCES

Amit, Y. (1996). Convergence properties of the Gibbs sampler for perturbations of Gaussians. *Ann. Statist.* **24** 122–140. MR1389883

Chen, R., Liu, J. S. and Wang, X. (2002). Convergence analyses and comparisons of Markov chain Monte Carlo algorithms in digital communications. *IEEE Trans. Signal Process.* **50** 255–270.

Diaconis, P., Khare, K. and Saloff-Coste, L. (2006). Gibbs sampling, exponential families and coupling. Preprint, Dept. Statistics, Stanford Univ.

Frigessi, A., Di Stefano, P., Hwang, C.-R. and Sheu, S.-J. (1993). Convergence rates of the Gibbs sampler, the Metropolis algorithm and other single-site updating dynamics. *J. Roy. Statist. Soc. Ser. B* **55** 205–219. MR1210432

Goodman, J. and Sokal, A. (1989). Multigrid Monte Carlo method conceptual foundations. *Phys. Rev. D* **40** 2035–2071.

Khare, K. and Zhou, H. (2008). Rates of convergence of some multivariate Markov chains with polynomial eigenfunctions. Technical report, Dept. Statistics, Stanford Univ.

Levine, R. A. (2005). A note on Markov chain Monte Carlo sweep strategies. *J. Stat. Comput. Simul.* **75** 253–262. MR2134639

Levine, R. A. and Casella, G. (2006). Optimizing random scan Gibbs samplers. *J. Multivariate Anal.* **97** 2071–2100. MR2301627

Levine, R. A., Yu, Z., Hanley, W. G. and Nitao, J. J. (2005). Implementing random scan Gibbs samplers. *Comput. Statist.* **20** 177–196. MR2162541

Liu, J., Wong, W. and Kong, A. (1995). Covariance structure and convergence rates of the Gibbs sampler with various scans. *J. Roy. Statist. Soc. Ser. B* **57** 157–169. MR1325382

Mira, A. (2001). Ordering and improving performance of Monte Carlo Markov chains. *Statist. Sci.* **16** 340–350. MR1888449

Peskun, P. H. (1973). Optimum Monte-Carlo sampling using Markov chains. *Biometrika* **60** 607–612. MR0362823

Roberts, G. O. and Sahu, S. K. (2001). Approximate predetermined convergence properties of the Gibbs sampler. *J. Comput. Graph. Statist.* **10** 216–229. MR1939698